\documentclass[a4paper,fleqn]{cas-dc}

\usepackage[numbers,compress]{natbib}

\newcommand*\EF{Eu(HCOO)$_3$}
\newcommand*\EFFA{Eu(HCOO)$_{3}\cdot$(HCONH$_2$)$_2$}

\def\tsc#1{\csdef{#1}{\textsc{\lowercase{#1}}\xspace}}
\tsc{WGM}
\tsc{QE}
\begin{document}
\let\WriteBookmarks\relax
\def\floatpagepagefraction{1}
\def\textpagefraction{.001}

\shorttitle{}    

\shortauthors{Z.W. Riedel et al.}  

\title [mode = title]{Synthesis of \EF\ and \EFFA\ crystals and observation of their $^5$D$_0$~$\rightarrow$~$^7$F$_0$ transition for quantum information systems}

\author[1,2]{Zachary W. Riedel}[orcid=0000-0001-5848-5520]
\author[2,3]{Donny R. Pearson}[suffix=Jr.]
\author[1,2]{Manohar H. Karigerasi}
\author[2]{Julio A.N.T. Soares}
\author[2,3,4]{Elizabeth A. Goldschmidt}
\ead{goldschm@illinois.edu}
\author[1,2]{Daniel P. Shoemaker}
\ead{dpshoema@illinois.edu}

\address[1]{Department of Materials Science and Engineering, University of Illinois at Urbana-Champaign, Urbana, Illinois 61801, United States}
\address[2]{Materials Research Laboratory, University of Illinois at Urbana-Champaign, Urbana, Illinois 61801, United States}
\address[3]{Department of Physics, University of Illinois at Urbana-Champaign, Urbana, Illinois 61801, United States}
\address[4]{US Army Research Laboratory, Adelphi, Maryland 20783, United States}

\begin{abstract}
Two stoichiometric metal-organic frameworks containing Eu$^{3+}$ cations are probed as candidates for photon-based quantum information storage. Synthesis procedures for growing 0.2~mm, rod-shaped \EF\ and 1-3~mm, rhombohedral \EFFA\ single crystals are presented with visible precipitation as soon as 1~h into heating for \EF\ and 24~h for \EFFA. Room temperature and 1.4~K photoluminescence measurements of the $^5$D$_0$~$\rightarrow$~$^7$F$_J$ transitions of Eu$^{3+}$ are analyzed for both compounds. Comparisons of peak width and intensity are discussed along with the notable first report for both of the $^5$D$_0$~$\rightarrow$~$^7$F$_0$ transition, the hyperfine structure of which has potential use in quantum memory applications. The air instability of \EFFA\ and the transformation of its photoluminescence properties are discussed.
\end{abstract}


% \begin{highlights}
% \item 
% \item 
% \item 
% \end{highlights}

% Keywords

\maketitle
\sloppy

% Main text
\section{Introduction}
Stoichiometric compounds with lanthanide cations in crystallographic sites can increase optical density for photon-based quantum information systems (QIS) while also reducing the optical transitions' inhomogeneous linewidths by avoiding local strain and disorder due to doping. Such a reduction is required to optically resolve the extremely long-lived hyperfine structure in these ions, such as the $^5$D$_0\rightarrow^7$F$_0$ transition in Eu$^{3+}$  \cite{ahlefeldt2009ligand, ahlefeldt2020quantum, zhong2015optically}. Achieving this limit of inhomogeneous linewidth smaller than the splitting between hyperfine levels is necessary to optically prepare the system in a particular state, which would enable long-lived and efficient quantum memory \cite{ahlefeldt2020quantum}, quantum information processing \cite{wesenberg2007scalable, kinos2021roadmap}, quantum transduction \cite{everts2019microwave}, etc. \cite{braggio2017axion} Currently, only EuCl$_{3}\cdot$6H$_2$O has been shown to achieve this inhomogeneous linewidth limit. Without isotopic purification, the $^5$D$_0$~$\rightarrow$~$^7$F$_0$ transition of Eu$^{3+}$ with its advantageous hyperfine levels has an inhomogeneous linewidth of 100 MHz in EuCl$_{3}\cdot$6H$_2$O \cite{ahlefeldt2009ligand}. Because of the proposed linear dependence of isotopic broadening on isotopes' relative mass difference, atomic concentration, and natural abundance \cite{ahlefeldt2016ultranarrow}, isotopic purification of EuCl$_{3}\cdot$6H$_2$O was performed for the Cl$^-$ ion. The resulting 99.67\% $^{35}$Cl purification reduced the linewidth to 25 MHz, smaller than the material's nearest-neighbor Eu$^{3+}$ interactions ($\sim$40~MHz) \cite{ahlefeldt2013precision} and small enough to resolve the hyperfine structure \cite{ahlefeldt2016ultranarrow}. But EuCl$_{3}\cdot$6H$_2$O crystals are highly hygroscopic and unstable under vacuum \cite{ahlefeldt2009ligand}. Therefore, selecting additional materials with low inhomogeneous linewidths and improved environmental stability is critical for advancing the study of stoichiometric rare-earth crystals for QIS. We identify and synthesize two candidate stoichiometric europium-containing crystals and investigate their structural and optical properties at room and cryogenic temperatures. We find that both candidates can be grown from moderately heated aqueous solutions in air, and exhibit the desired $^7$F$_0$~$\rightarrow$~$^5$D$_0$ transition with typical optical lifetime values.

\begin{figure}
    \centering
    \includegraphics[width=\columnwidth]{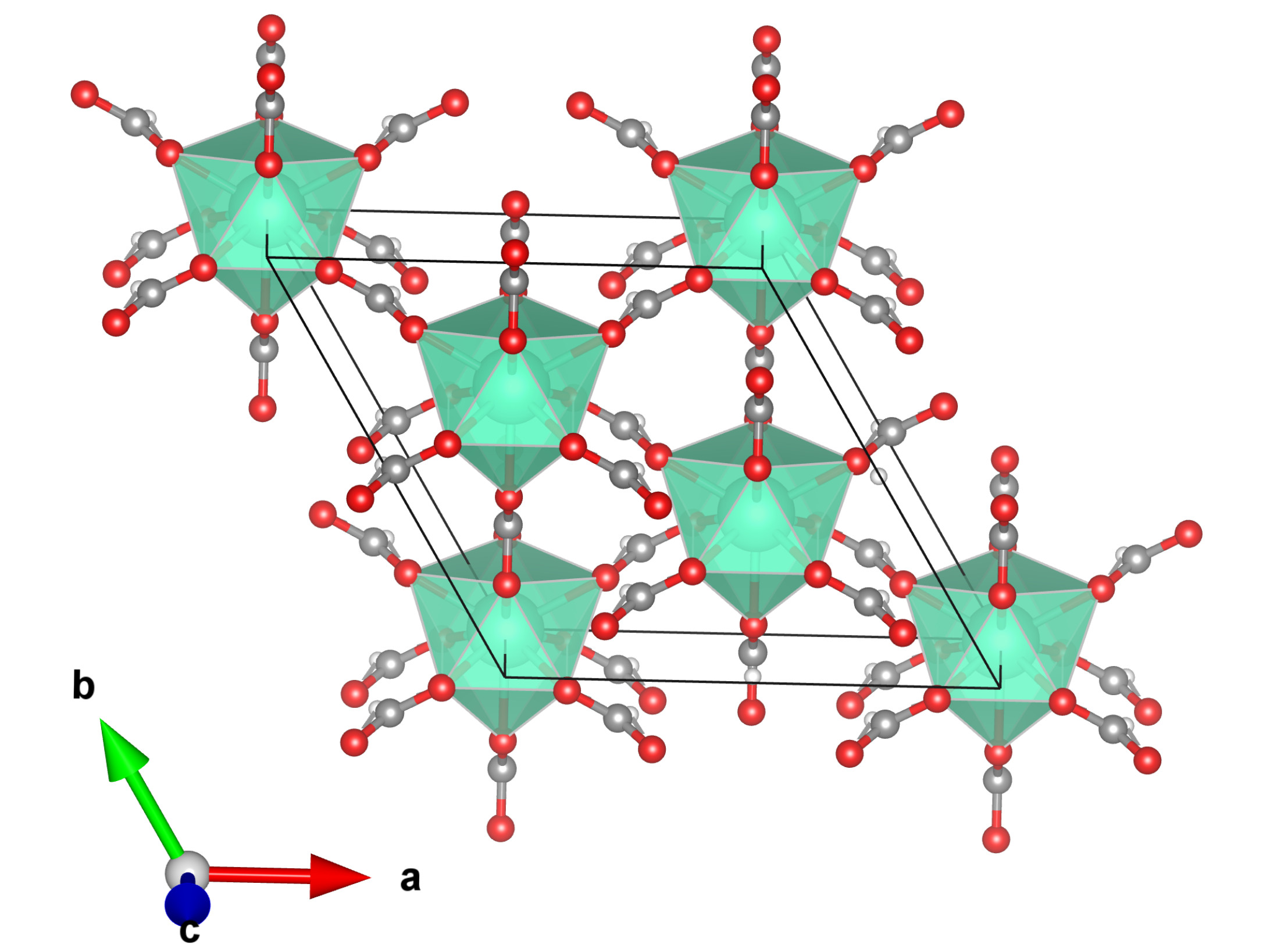}
    \caption{Crystal structure of \EF\ ($R3m$, $a$~=~$b$~=~10.4972~\AA, $c$~=~4.0011~\AA, $\alpha$~=~$\beta$~=~90$^{\circ}$, $\gamma$~=~120$^{\circ}$) [Eu: teal, H: white, C: gray, O: red]}
    \label{fig:EFstruct}
\end{figure}

\begin{figure}
    \centering
    \includegraphics[width=\columnwidth]{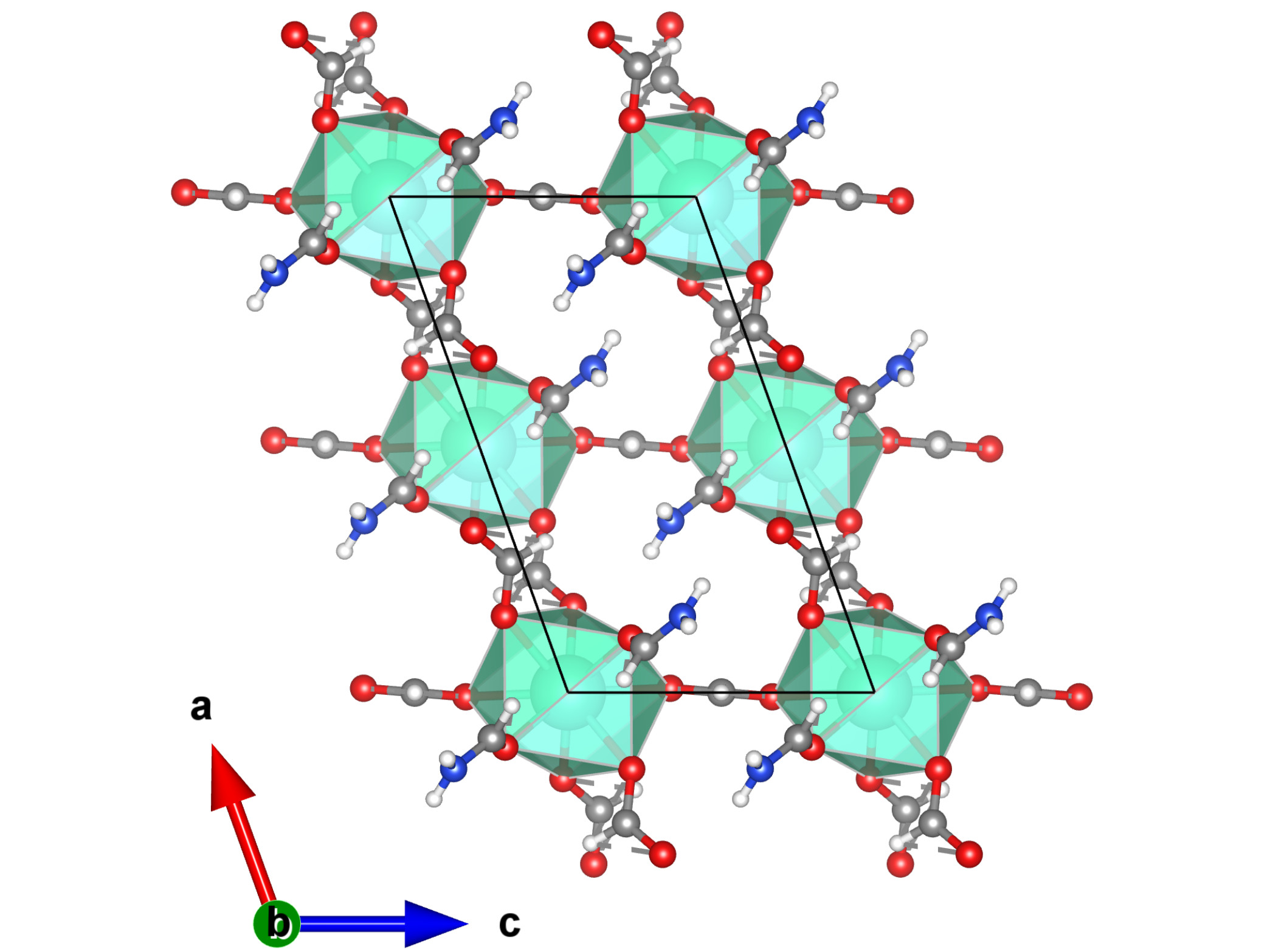}
    \caption{Crystal structure of \EFFA\ ($C2$, $a$~=~11.4164~\AA, $b$~=~7.1985~\AA, $c$~=~6.6405~\AA, $\beta$~=~109.829$^{\circ}$) [Eu: teal, H: white, C: gray, O: red, N: blue]}
    \label{fig:EFFAstruct}
\end{figure}

Europium formate \EF\ [``\textbf{EF}"] and europium formate formamide \EFFA\ [``\textbf{EF$\cdot$FA}"] are two promising candidates for QIS applications. Both compounds can be obtained by solution precipitation \cite{samarasekere2015reactions,zhao2019facile} and have  published crystal structures, shown in Figures~\ref{fig:EFstruct} and \ref{fig:EFFAstruct} (EF: CSD Entry ZZZVME01 \cite{groom2016cambridge}, EF$\cdot$FA: ICSD Entry 251671 \cite{belsky2002new}). A procedure for reliably producing mm-scale crystals that are environmentally stable is necessary for practical use in a quantum memory device and for the possibility of future photonic integration. 


EF consists of Eu$^{3+}$ cations nine-fold coordinated to formate groups that bridge them. EF$\cdot$FA on the other hand has eight-fold coordinated Eu$^{3+}$ cations with six Eu$^{3+}$-bridging formate groups and two terminal formamide groups. The compounds' non-europium elements have a dominant naturally abundant isotope ($^1$H: 99.99\%, $^{12}$C: 98.89\%, $^{16}$O: 99.76\%, $^{14}$N: 99.64\%), reducing the need for isotopic purification to promote a narrow optical linewidth. Still, deuteration of the compounds would be expected to improve quantum efficiency and transition coherence time \cite{ahlefeldt2013optical}. Both have relatively large Eu$^{3+}$ separation within the material ($\geq$4.001~\AA\ in EF and $\geq$6.640~\AA\ in EF$\cdot$FA), which is desirable to prevent interactions between the Eu$^{3+}$ optical centers. 

\section{Experimental Procedures}
\subsection{Synthesis}
Numerous clear, faceted EF and EF$\cdot$FA crystals were produced from a solution of formamide and de-ionized water. For both, Eu(NO$_3$)$_{3}\cdot$5H$_2$O (Sigma-Aldrich, 99.9\% REO) was added to a 7~mL glass vial along with a 3.5~mL solution of 98\% formamide HCONH$_2$ (Alfa Aesar, 99\%) and 2\% de-ionized H$_2$O, by volume. The Eu(NO$_3$)$_{3}\cdot$5H$_2$O concentration was 0.05~M for EF and 0.075~M for EF$\cdot$FA. The solutions were gently swirled and inverted until the powder was dissolved. The uncovered vials were heated for 48~h on a hot plate with the solution temperature at 125$^{\circ}$C for EF and 65$^{\circ}$C for EF$\cdot$FA. A camera periodically captured images of the crystal precipitation (Supplementary Material). Gravity filtration was used to separate the crystals from the reaction solution, followed by rinsing with ethanol and drying at 80$^{\circ}$C for 20~min.  

\subsection{Characterization}
Scanning electron microscopy images were collected with a JEOL 6060LV SEM. A Bruker D8 diffractometer equipped with a capillary mount was used for powder X-ray diffraction. Confocal room temperature photoluminescence measurements on individual crystals were obtained with 532~nm excitation on a Nanophoton Raman 11 microscope. A 405~nm laser and a SpectraPro 300i spectrograph were used to collect 1.4~K and 298~K photoluminescence spectra. Samples were mounted vertically with Cu tape on a Cu plate in a Janis 10DT liquid He bath cryostat. Data were collected for a single EF$\cdot$FA crystal and for a cluster of EF ones. For lifetime measurements, the fluorescent emission was collected with a single-photon avalanche diode and accumulated into a histogram of 1500 time bins with a bin width of 10~$\mu$s. The cryogenic EF$\cdot$FA data was collected using 66.3~$\mu$W of excitation power with a 300~$\mu$s pulse duration and was integrated for 6~min. The cryogenic EF data was collected using 6.58~$\mu$W excitation power with a 100~$\mu$s pulse duration and was integrated for 2.5~min. Inductively coupled plasma mass spectrometry (ICP-MS) was performed with a PerkinElmer NexION 350D using standards with concentrations as low as 0.5~ppm.

\section{Results and Discussion}
\subsection{Crystal Growth}
\begin{figure}
    \centering\includegraphics[width=0.491\columnwidth]{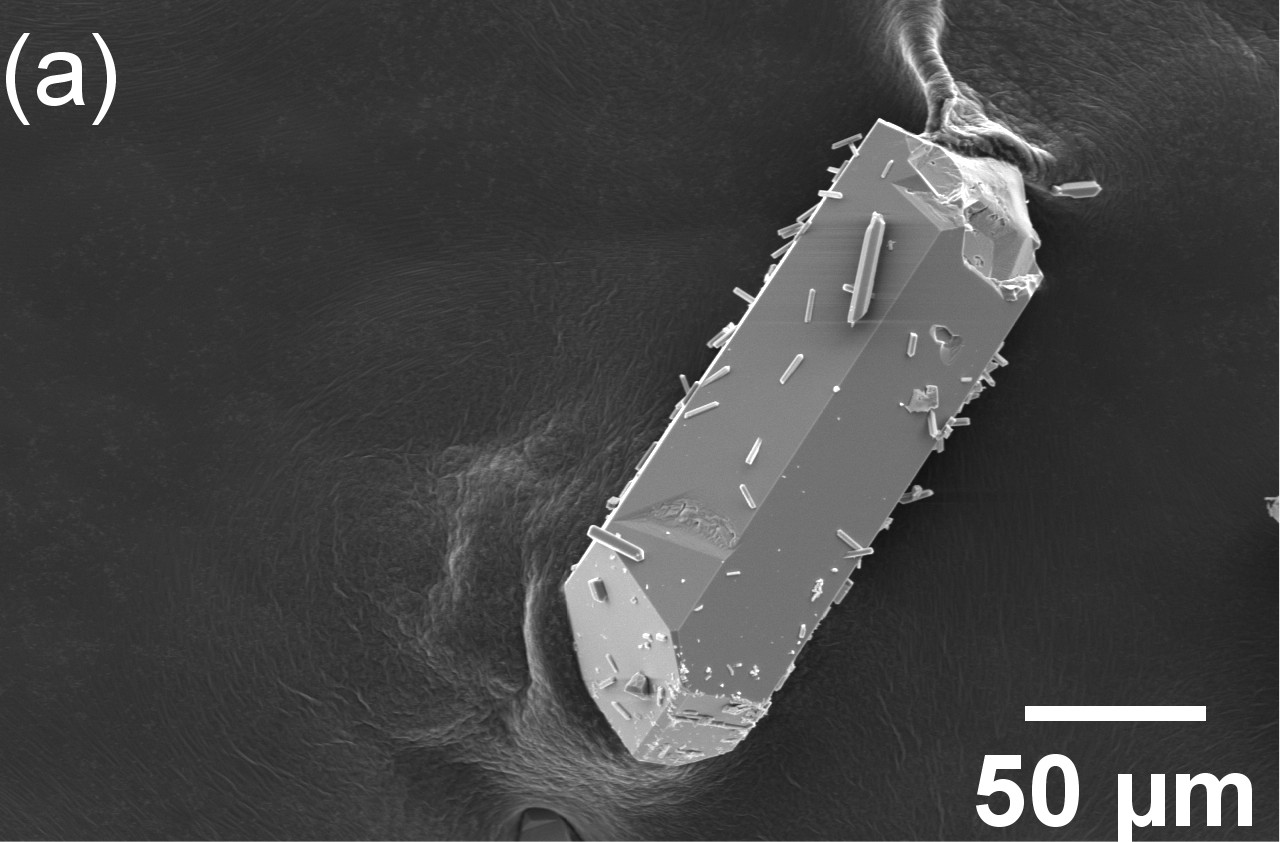}
    \hfill
    \centering\includegraphics[width=0.491\columnwidth]{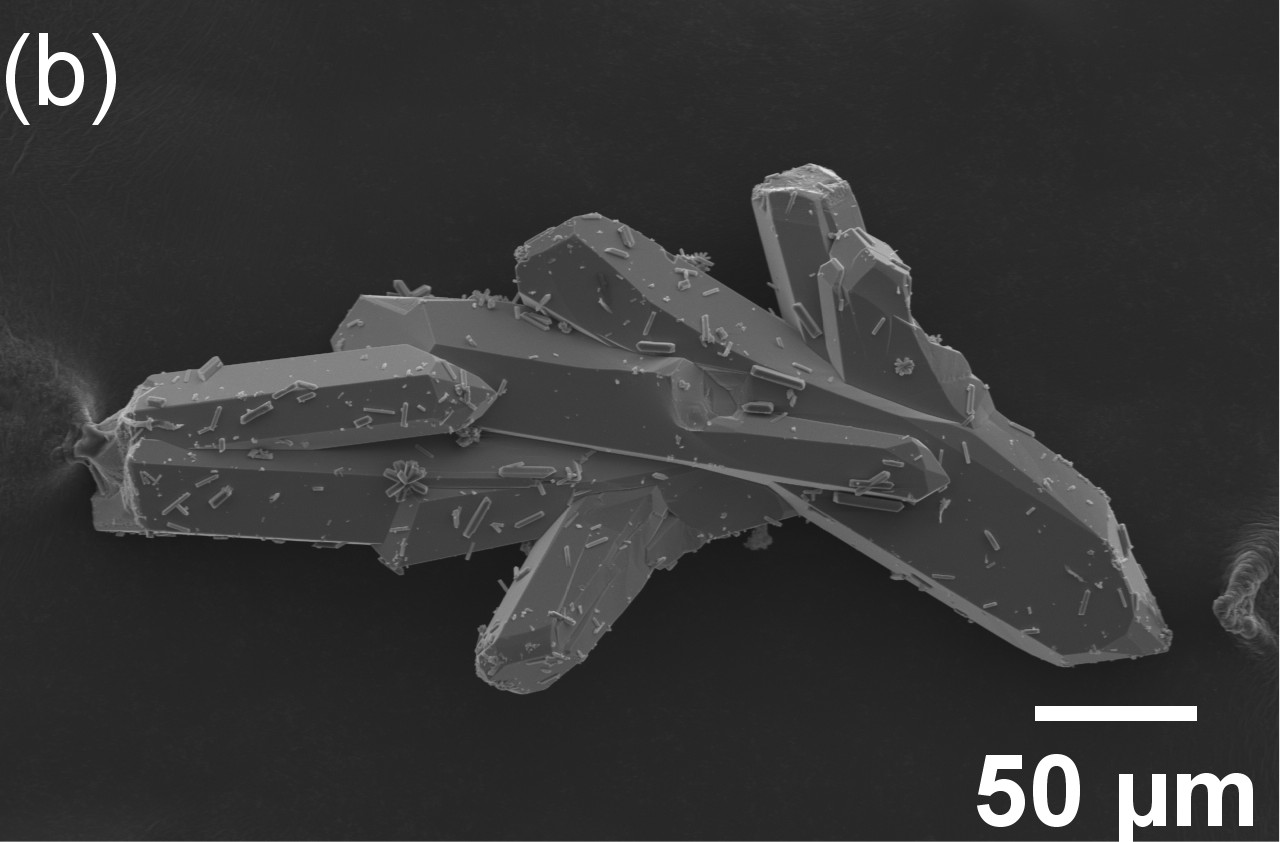}
    \hfill
    \centering\includegraphics[width=0.491\columnwidth]{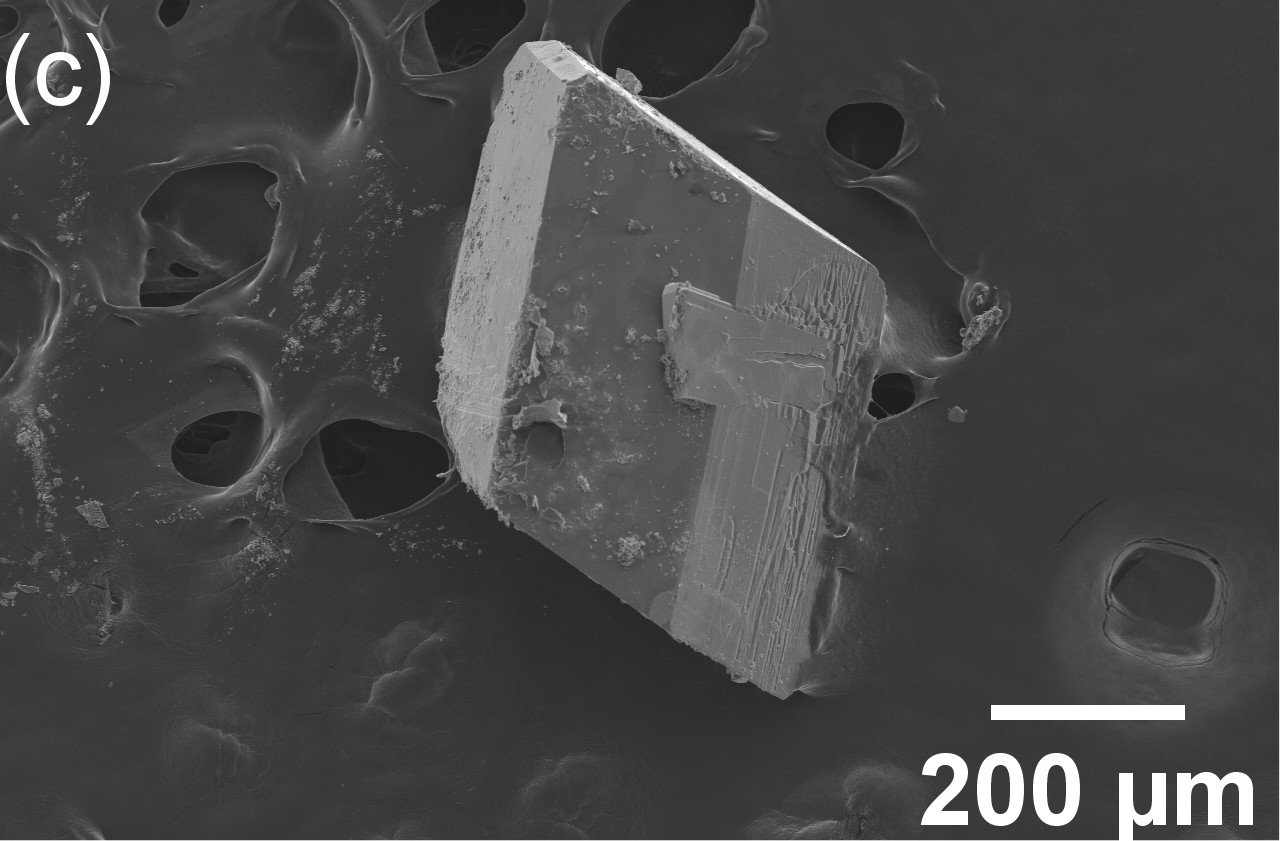}
    \hfill
    \centering\includegraphics[width=0.491\columnwidth]{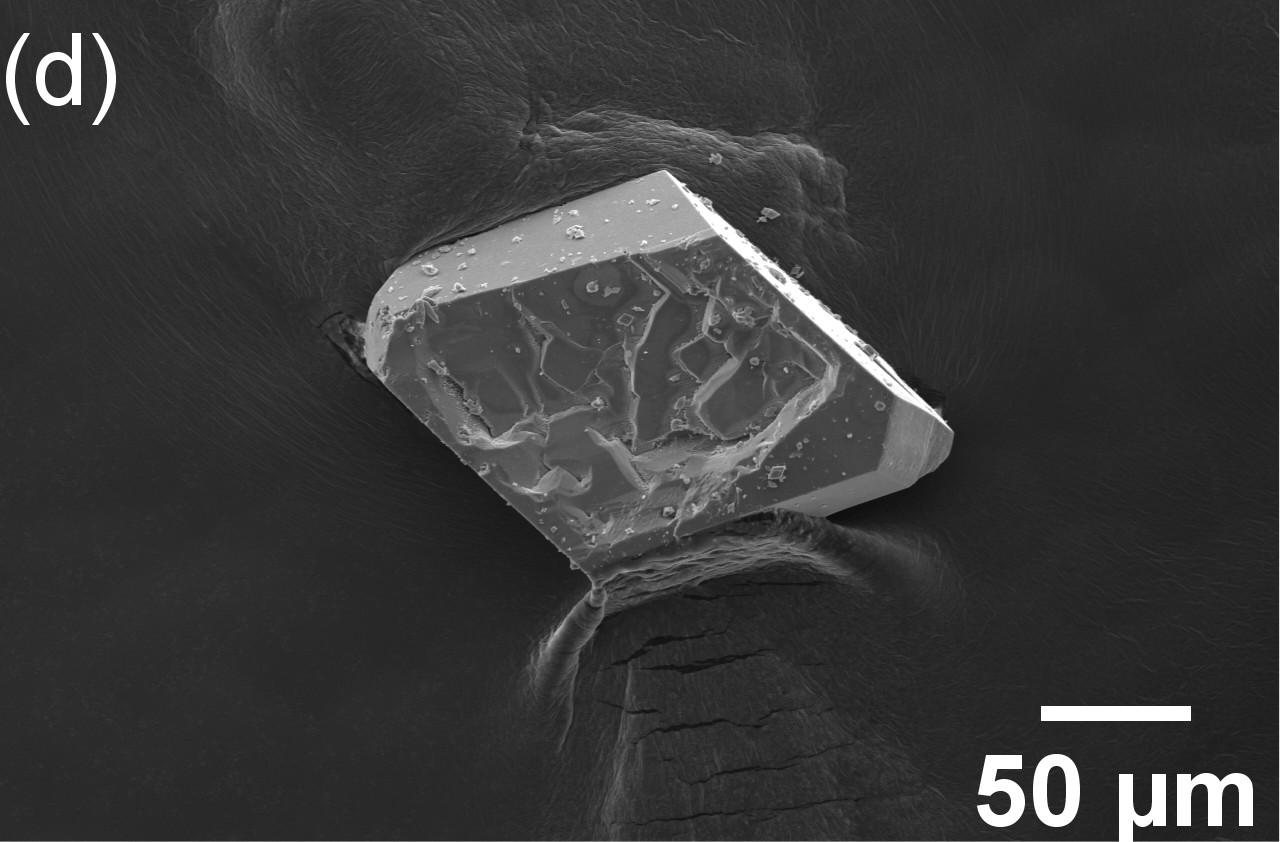}
    \caption{(a) EF rod crystal with smaller EF crystals dotting the surface (b) Cluster of EF crystals that have grown off each other (c,d) EF$\cdot$FA rhombohedral crystals with surface defects}
    \label{fig:SEM}
\end{figure}

Many clusters of EF crystals formed with a Eu content yield of 71.1\% after 48~h. Time lapse video showed visible precipitation from the solution beginning after only 1~h of heating without the presence of a seed crystal. For EF$\cdot$FA, rhombohedral crystals formed with a Eu content yield of 53.0\% after 48~h. Visible precipitation began around 24~h without a seed crystal. EF and EF$\cdot$FA crystals appeared to form throughout the solution, with some dropping to the bottom of the vial and some being stuck on the side of the vial at the liquid-air interface. SEM images showed that the EF crystals had nice facets but were frequently fused together and had smaller crystals sitting on their surface; imaging of EF$\cdot$FA crystals consistently revealed surface defects (Figure~\ref{fig:SEM}). The lengthier growth process of the EF$\cdot$FA crystals at a lower temperature promoted the formation of larger crystals with sides around 1~mm long (Figure~\ref{fig:DecompFilm}). Moreover, introducing a seed crystal to the same initial solution led to the formation of a EF$\cdot$FA crystal with 3~mm sides along with additional 1~mm scale crystals. Seeding did not lead to any obvious differences for EF.

The faster growth of smaller EF crystals can be explained kinetically. The hydrolysis of formamide produces ammonia and formic acid, thus providing the formate ions incorporated in the two structures \cite{samarasekere2015reactions}. Three pathways for the hydrolysis of formamide have been proposed, acid catalyzed, base catalyzed, and neutral water catalyzed. The neutral catalyzed pathway, although advocated for by experimental fitting \cite{slebocka2002aspects} and a computational study \cite{gorb2005mechanism}, has been rejected by some on the basis of its high calculated activation energy \cite{almerindo2007ab}. Still, Miyakawa et al. studied the hydrolysis kinetics and provided an experimental fit to the first-order rate constant for formamide hydrolysis based on H$^+$ and OH$^-$ concentration \cite{miyakawa2002cold}. pH paper measurements at the beginning and end of the crystal growths indicate that the reaction solutions remain approximately neutral during the growth, and the water content of the two solutions, based on added DI H$_2$O and Eu(NO$_3$)$_{3}\cdot$5H$_2$O concentration, are similar (4.8~mmol H$_2$O for EF and 5.2~mmol for EF$\cdot$FA). Utilizing the Miyakawa equations, at pH~7, the rate of hydrolysis in the 125$^{\circ}$C EF solution (2.4$\times$10$^{-7}$~s$^{-1}$) should be 38 times larger than in the 65$^{\circ}$C EF$\cdot$FA solution (6.3$\times$10$^{-9}$~s$^{-1}$). This indicates that formate ions should form much more quickly in the EF solution, leading to the observed faster growth.

\subsection{Room Temperature Photoluminescence}
\begin{figure}
    \centering
    \includegraphics[width=\columnwidth]{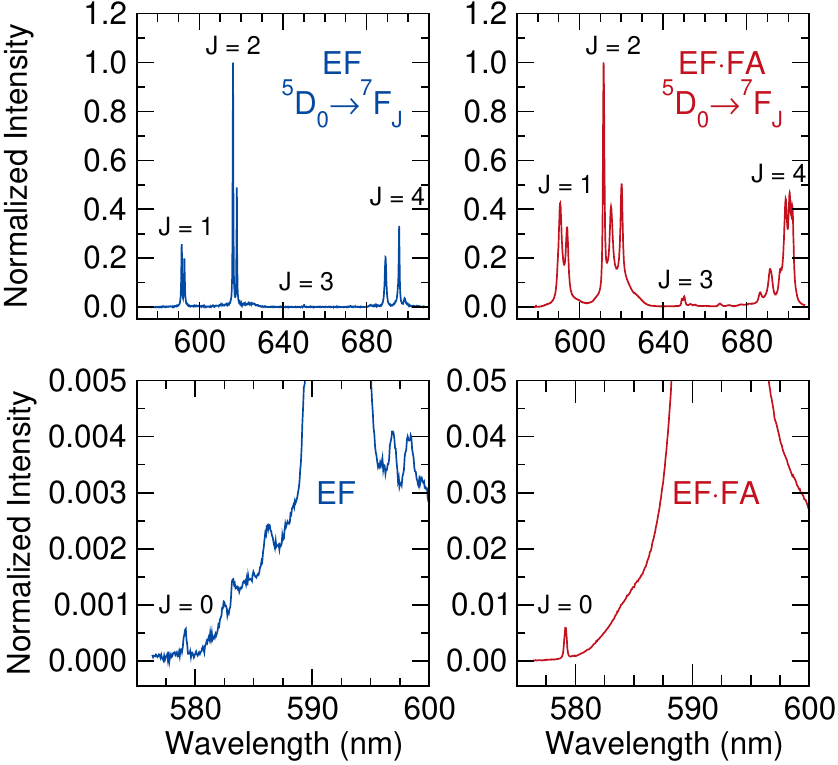}
    \caption{Photoluminescence emission spectra of EF (left column) and EF$\cdot$FA (right column) recorded with an excitation wavelength of 532~nm.}
    \label{fig:RTPL}
\end{figure}

% noisier in EF
The room temperature emission spectra of EF and EF$\cdot$FA are shown in Figure~\ref{fig:RTPL}. Both show peaks for the $^5$D$_0$~$\rightarrow$~$^7$F$_J$ set of transitions. Comparing the spectra with literature values \cite{samarasekere2015reactions, zhao2019facile, binnemans2015interpretation, sohn2014photoluminescence}, the EF transitions centered at 579, 592, 616, 650, and 696~nm are assigned to $J$~=~0, 1, 2, 3, 4, respectively, and for EF$\cdot$FA, the same transitions are centered at 579, 591, 612, 650, and 701~nm. This is the first observation of the $^5$D$_0$~$\rightarrow$~$^7$F$_0$ transition for either material.

The noticeably sharper transitions for EF indicate phonon broadening effects associated with the differing ligands. Although many other Eu$^{3+}$-containing systems have demonstrated large broadening effects due to a higher degree of crystallinity or a lower number of defects within the sample \cite{d2021morphology,driesen2007eu,mohanty2006light,munoz2009structural,wu2010crystallinity}, those explanations do not hold here as even the EF$\cdot$FA crystals that decompose into EF experience drastic transition sharpening (Supplementary Material). Accompanying this sharpening is the appearance of additional peaks around the $^5$D$_0$~$\rightarrow$~$^7$F$_1$ transition of EF. These peaks, presumably, arise from rare-earth impurities in the precursor. Although Nd$^{3+}$ has absorption spectrum $^4$I$_{9/2}$~$\rightarrow$~$^4$G$_{5/2}$~+~$^2$G$_{7/2}$ peaks that overlap with the $^5$D$_0$~$\rightarrow$~$^7$F$_1$ transition of Eu$^{3+}$ \cite{harmer1969fluorescence,jose2021effective,manasa2020optical} and Gd$^{3+}$ has emission peaks for $^6$G$_J$~$\rightarrow$~$^6$P$_J$ transitions in the same region \cite{zhong2010luminescence}, neither cation was detected using ICP-MS. The extra peaks also appear in samples decomposing from EF$\cdot$FA to EF, but in that case, they are lower in intensity than the $J$~=~0 peak (Supplementary Material).

The $^5$D$_0$~$\rightarrow$~$^7$F$_1$ and $^5$D$_0$~$\rightarrow$~$^7$F$_2$ transitions are frequently used to compare local coordination environments of Eu$^{3+}$. The intensity of the induced electric dipole transition $^5$D$_0$~$\rightarrow$~$^7$F$_2$ is strongly dependent on the environment of Eu$^{3+}$ while the magnetic dipole transition $^5$D$_0$~$\rightarrow$~$^7$F$_1$ is mostly independent of it \cite{binnemans2015interpretation}. As a result, an increase in the ratio of integrated intensities, I($^5$D$_0$~$\rightarrow$~$^7$F$_2$)/I($^5$D$_0$~$\rightarrow$~$^7$F$_1$), can indicate distortions from centrosymmetric geometry for Eu$^{3+}$ \cite{tanner2013some}. Both experimental EF and EF$\cdot$FA structures contain Eu$^{3+}$ ions in non-centrosymmetric sites ($C_{3v}$ symmetry in EF and $C_2$ in EF$\cdot$FA), so large values ($\sim$2-4 \cite{gao2010enhanced,prasad2013bi,volanti2009role}) of the parameter are expected. These values are shown in the $J$~=~2 column of Table~\ref{tab:PkIntIntensity}. The site in EF, with its higher symmetry, shows a larger ratio than that of the ion in EF$\cdot$FA. The EF$\cdot$FA value of 2.0 exceeds that reported previously of 1.75 \cite{samarasekere2015reactions}. Although these values were obtained at different excitation wavelengths, they can be compared since the ratio's excitation wavelength dependence is associated with site selective excitation arising from multiple Eu$^{3+}$ sites in the material \cite{kolesnikov2018asymmetry,rambabu2013luminescence,szczeszak2011spectroscopic}, and EF$\cdot$FA only has one Eu$^{3+}$ crystallographic site.

\begin{table}
\small
\centering 
\caption{\label{tab:PkIntIntensity} Room temperature integrated intensities of $^5$D$_0$~$\rightarrow$~$^7$F$_J$ peaks normalized to the magnetic dipole $J$~=~1 transition with sample standard deviations in parentheses} 
\resizebox{\columnwidth}{!}{\begin{tabular}{l l l l l l}
\hline 
Compound & $J=0$ & $J=1$ & $J=2$ & $J=3$ & $J=4$
\\
\hline
EF & 0.0011 & 1.0 & 3.2 & 0.019 & 2.0 \\
 & (0.00026) & (-) & (0.14) & (0.0012) & (0.19) \\
EF$\cdot$FA & 0.0010 & 1.0 & 2.0 & 0.066 & 1.6 \\
 & (0.00018) & (-) & (0.17) & (0.028) & (0.48)
\end{tabular}}
\end{table}

The appearance of the $^5$D$_0$~$\rightarrow$~$^7$F$_3$ transition for the two compounds is also notable. This transition is forbidden by the Judd-Ofelt rules for magnetic and electric dipole transitions, but crystal field induced \textit{J}-mixing can lead to the transition's presence in the spectrum \cite{binnemans2015interpretation,chen2005standard,lowther1974spectroscopic}. For several compounds, the $^7$F$_3$ state has been shown to mix with the $^7$F$_2$ and $^7$F$_4$ states \cite{mohanty2006light,lowther1974spectroscopic}.

The hypersensitive $^5$D$_0$~$\rightarrow$~$^7$F$_0$ transition, although also forbidden, is allowed for Eu$^{3+}$ ions with non-cubic, non-centrosymmetric site symmetries \cite{tanner2013some,binnemans1996application}. Therefore, it can, by point symmetry, appear for EF and EF$\cdot$FA, but it has not been reported previously for either. In the spectra normalized to the maximum $^5$D$_0$~$\rightarrow$~$^7$F$_2$ peak height (Figure~\ref{fig:RTPL}), the EF$\cdot$FA $^5$D$_0$~$\rightarrow$~$^7$F$_0$ peak appears approximately an order of magnitude larger than the one in EF. However, when more appropriately compared through a normalization to the integrated intensity of the environment independent $^5$D$_0$~$\rightarrow$~$^7$F$_1$ transition, the intensities of the $^5$D$_0$~$\rightarrow$~$^7$F$_0$ transitions in the two compounds are remarkably similar (Table~\ref{tab:PkIntIntensity}).

\subsection{Cryogenic Photoluminescence}

\begin{figure}
    \centering
    \includegraphics[width=\columnwidth]{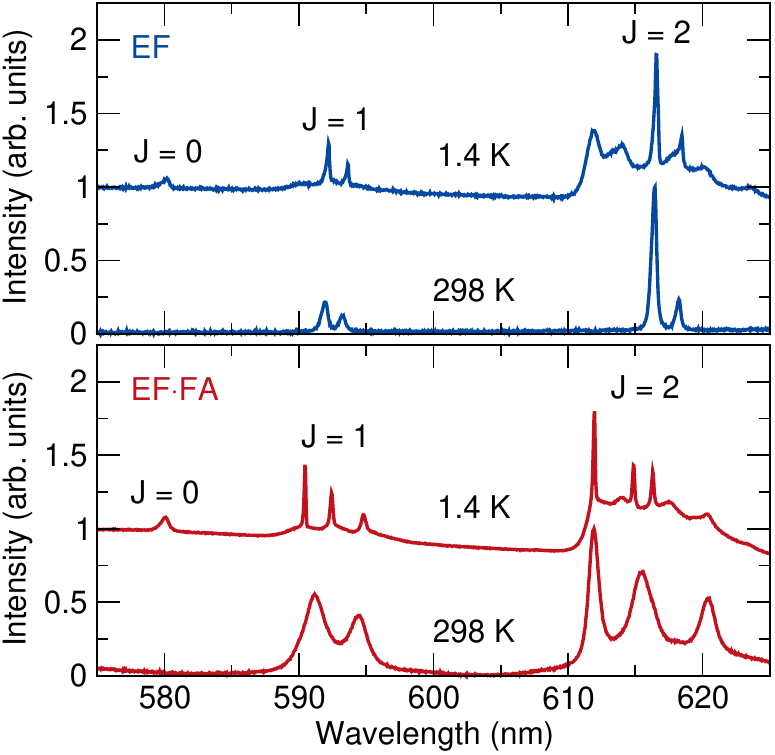}
    \caption{Photoluminescence data collected at 1.4~K using an excitation wavelength of 405~nm with the $^5$D$_0$~$\rightarrow$~$^7$F$_J$ transitions labelled. The $J$~=~0 peak appears for both while it is not visible at the same collection times at room temperature.}
    \label{fig:cryoPL}
\end{figure}

Photoluminescence emission spectra for both compounds were collected at 1.4~K in order to increase the intensity of the $^5$D$_0$~$\rightarrow$~$^7$F$_J$ transitions and to eliminate phonon broadening. Figure~\ref{fig:cryoPL} shows these spectra along with room temperature spectra collected with the same mounting setup for comparison. Even at low laser power, the $J$~=~0 transition is clearly seen for both compounds. The integrated intensity of the transition's peak relative to those of $J$~=~1 is now 0.11 for EF and EF$\cdot$FA, 100x larger than in the room temperature data in Table~\ref{tab:PkIntIntensity}.

The EF$\cdot$FA peaks are noticeably less broad than at room temperature and have comparable widths to those of EF. This further confirms that EF$\cdot$FA's broad room temperature spectrum is a product of phonon effects, which are mitigated at 1.4~K. Moreover, the improved resolution allows for the observation of additional splittings in the EF$\cdot$FA $J$~=~1 and $J$~=~2 peaks. The splitting of the $J$~=~1 peak is indicative of the site symmetry of the Eu$^{3+}$ ion. The transition in EF has two components, as could be seen at room temperature, and the expected three components of the transition for EF$\cdot$FA \cite{bunzli2011} are now resolved at 1.4~K. 

Both compounds experience peak shifts to longer wavelengths at 1.4~K. In other Eu$^{3+}$ systems, as temperatures lower, $^5$D$_0$~$\rightarrow$~$^7$F$_J$ transition red-shifts are typically noted \cite{arashi1982diamond,hellwege1951spektrum,kusama1976line}. Here, for the most prominent $J$~=~2 peak for each compound in the 405~nm data (Figure~\ref{fig:cryoPL}), the shifts are +0.09~nm for EF and +0.04~nm for EF$\cdot$FA, quite small for both. To compare the $J$~=~0 peak positions, observed in the 532~nm room temperature and the 405~nm 1.4~K spectra, it must be noted that the 405~nm room temperature spectrum has a peak shift of +0.30~nm compared to the 532~nm room temperature data. This is likely due to a difference in the calibration techniques for the two instrument setups and needs to be accounted for when comparing the $J$~=~0 peak positions. Since the maxima of the $J$~=~0 peaks in the 532~nm room temperature data are at 579.19~nm for EF and 579.13~nm for EF$\cdot$FA and are at 580.20~nm (+1.01~nm) for EF and 580.11~nm (+0.98~nm) for EF$\cdot$FA in the 405~nm 1.4~K data, the corrected shifts are roughly +0.71~nm for EF and +0.68~nm for EF$\cdot$FA. The peak shifts, therefore, vary between the Eu$^{3+}$ transitions but are fairly consistent between the two compounds.

\begin{figure*}
    \centering
        \centering\includegraphics[width=0.96\columnwidth]{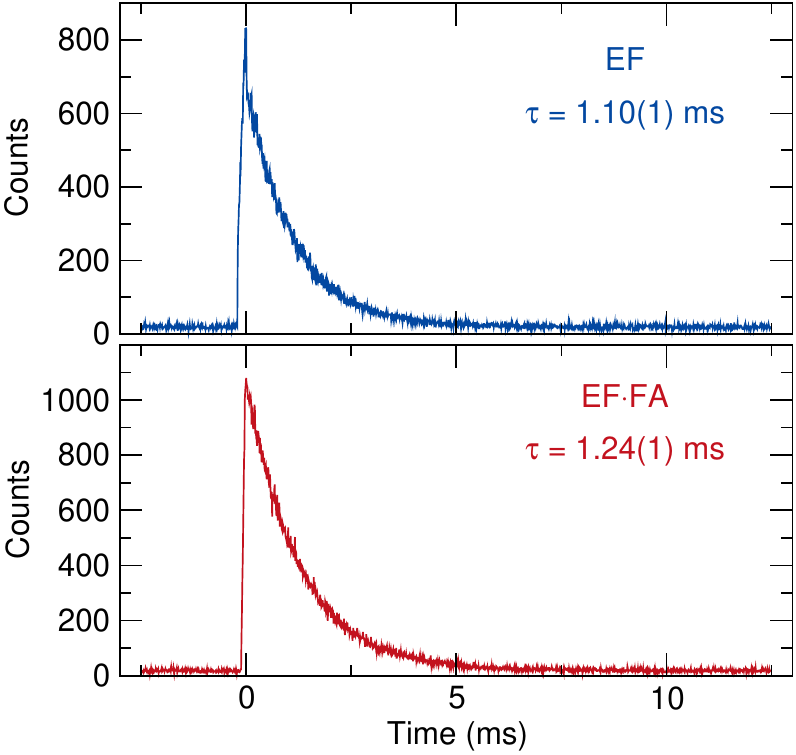}
    \hspace{0.2cm}
        \centering\includegraphics[width=0.96\columnwidth]{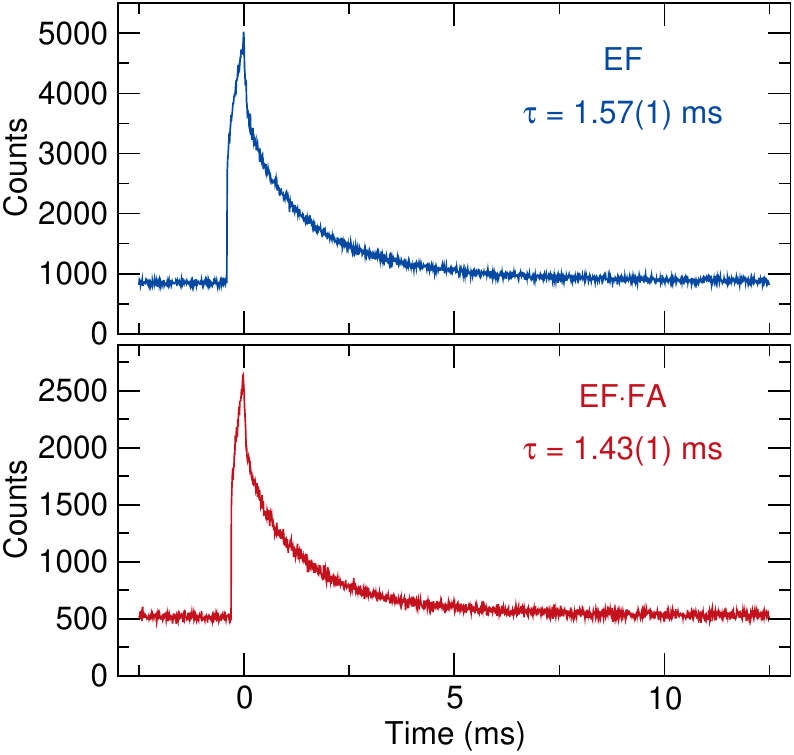}
        \caption{(left) Room temperature photoluminescence decay and (right) 1.4~K photoluminescence decay using an excitation wavelength of 405~nm.}
        \label{fig:Lifetimes}
\end{figure*}

\subsection{Photoluminescence Lifetime}
While a room temperature photoluminescence lifetime for EF's $^5$D$_0$ level has previously been reported as 1.06~ms \cite{zhao2019facile}, the cryogenic value has not been reported, and no lifetime measurements have been reported for EF$\cdot$FA. The room temperature lifetime of EuCl$_{3}\cdot$6H$_2$O, the other stoichoimetric rare-earth crystal studied for QIS, is 0.130~ms \cite{freeman1964effect}. Lifetime plots are shown in Figure~\ref{fig:Lifetimes}.

Photoluminescence decay was observed using 405~nm excitation pulses where the pulse duration, $t_{exc}$, is short compared to the observed lifetime, $\tau$. The excitation pulses were transmitted to the samples through a short-pass dichroic mirror. Photoluminescence emission from the materials was then collected from the dichroic mirror reflection, passed through a pair of 550~nm long-pass filters, and sent to a single-photon avalanche diode (SPAD) via coupling into a multi-mode fiber. The collected emission was then averaged over many cycles. Further details can be found in the Supplementary Material. The decay for both materials was fitted to a single-exponential with a vertical offset to account for background and dark counts.

The room temperature lifetime of EF, 1.10$\pm$0.01~ms, is similar to that reported previously, and the room temperature lifetime of EF$\cdot$FA is 1.24$\pm$0.01~ms. The cryogenic lifetimes for EF (1.57$\pm$0.01~ms) and for EF$\cdot$FA (1.43$\pm$0.01~ms) exceed by an order of magnitude the lifetime observed for EuCl$_{3}\cdot$6H$_2$O below 4~K (0.116$\pm$0.001~ms) but are shorter than its deuterium analog (2.6~ms) \cite{ahlefeldt2013optical}. Both compounds show an increase in lifetime with decreasing temperature. An increase is common in Eu$^{3+}$ compounds \cite{berry1996temperature,katumo2020smartphone,kitagawa2021site,murakami2000photoluminescence,nikolic2013temperature,sevic2019yvo4,west1992comparison}, but this trend is occasionally not consistent \cite{deng2018eu3+,grigorjevaite2016luminescence}, notably in the case of EuCl$_{3}\cdot$6H$_2$O \cite{ahlefeldt2013optical,freeman1964effect}.

\subsection{Phase Stability}
\begin{figure}
    \centering
        \centering\includegraphics[width=0.48\columnwidth]{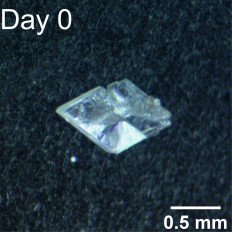}
    \hfill
        \centering\includegraphics[width=0.48\columnwidth]{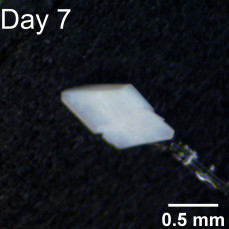}
        \caption{(left) EF$\cdot$FA crystal with $\sim$1~mm sides shown after growth and (right) after a week in air. The crystals form a film of EF on the surface before transforming further into an opaque crystal.}
        \label{fig:DecompFilm}
\end{figure}

Both synthesis procedures produce phase pure crystals as confirmed by powder X-ray diffraction. Over time, however, with exposure to air, EF$\cdot$FA crystals form a film of EF on their surface, giving them a cloudy appearance before eventually appearing opaque (Figure~\ref{fig:DecompFilm}). This transformation is also confirmed by single crystal X-ray diffraction, showing diffraction spots corresponding to EF$\cdot$FA and powder rings corresponding to EF (Supplementary Material).

\begin{figure}
    \centering
    \includegraphics[width=\columnwidth]{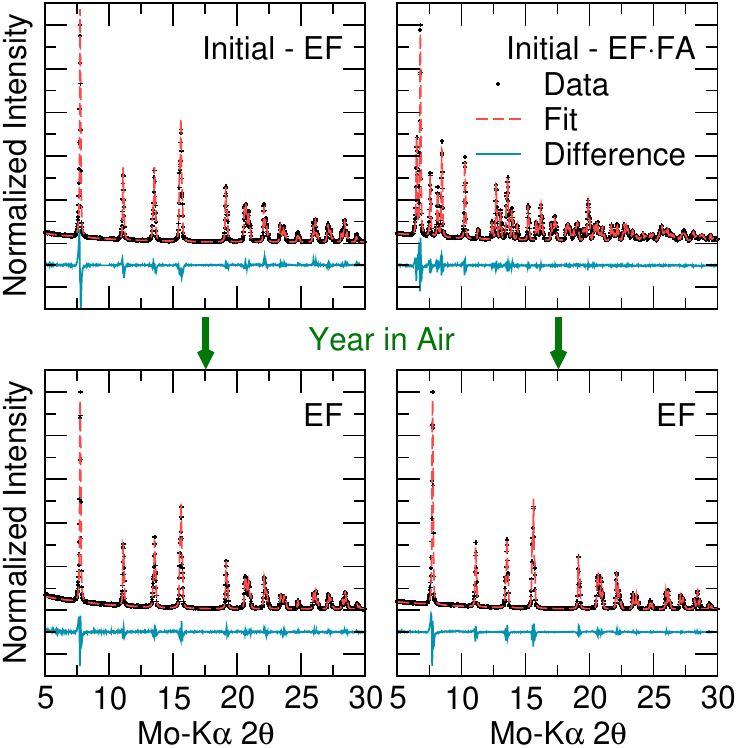}
    \caption{Powder X-ray diffraction data demonstrating the complete decomposition of EF$\cdot$FA (right column) into EF and the stability of EF (left column) over time when stored in air.}
    \label{fig:DecompXRD}
\end{figure}

The complete structural transformation to EF is not immediate, but high humidity will cause significant changes in the material's optical properties after only a day as the room temperature photoluminescence pattern shifts quickly from that of EF$\cdot$FA to that of EF. The decomposition of five crystals was monitored during storage in air at 50-55\% relative humidity. In particular, the distinguishing I($^5$D$_0$~$\rightarrow$~$^7$F$_2$)/I($^5$D$_0$~$\rightarrow$~$^7$F$_1$) ratio, noted before for its sensitivity to crystal symmetry, was tracked as a measure of spectrum transformation. Full transformation of the emission spectrum occurred in a week. Plots and images detailing this transformation are provided in the Supplementary Material. Powder X-ray diffraction of crushed EF$\cdot$FA crystals stored in air showed a full phase transformation into EF after a year (Figure~\ref{fig:DecompXRD}). The storage of EF$\cdot$FA in a desiccator or inert atmosphere prevents this decomposition into EF. On the other hand, EF remains stable over the same period of time.
 
\section{Conclusions}
Quick synthesis procedures were developed for forming faceted 0.2~mm crystals of EF and 1-3~mm crystals of EF$\cdot$FA. The first reported recording of the $^5$D$_0$~$\rightarrow$~$^7$F$_0$ transition was shown for EF and EF$\cdot$FA. Emission spectra at 1.4~K revealed a one hundred fold increase in the relative intensity of the transition and a reduction in phonon broadening effects for EF$\cdot$FA. The $^5$D$_0$~$\rightarrow$~$^7$F$_0$ peak for both materials shifted by roughly +0.7~nm upon cooling. The materials exhibit lifetimes of 1.57~ms and 1.43~ms at 1.4~K for EF and EF$\cdot$FA, respectively. The decomposition of the EF$\cdot$FA crystals into EF due to air exposure was shown and eliminated by storage in a desiccator or inert atmosphere. EF remains stable in air over time. Photoluminescence excitation and Raman heterodyne spectroscopy will be necessary to probe the materials' $^5$D$_0$~$\rightarrow$~$^7$F$_0$ inhomogeneous width and hyperfine structures at cryogenic temperatures.

\section{Acknowledgements}
Research for this work was carried out in part in the Materials Research Laboratory Central Research Facilities, University of Illinois. This work is supported by the U.S. Department of Energy, Office of Science, National Quantum Information Science Research Centers.
Z.W.R.\ was supported by a DIGI-MAT fellowship from the NSF DGE program, Award No.\ 1922758. 

\bibliographystyle{elsarticle-num}

\bibliography{main.bib}

\end{document}

% \maketitle

\begin{center}
\Large 
\textbf{Synthesis of \EF\ and \EFFA\ crystals and observation of their $^5$D$_0$~$\rightarrow$~$^7$F$_0$ transition for quantum information systems}\\
\vspace{1em}
Supplementary Material\\
\vspace{1em}
\normalsize
Zachary W. Riedel, Donny R. Pearson Jr., Manohar H. Karigerasi, Julio A.N.T. Soares, Elizabeth A. Goldschmidt, Daniel P. Shoemaker
\end{center}

\vspace{2em}

\begin{figure}
    \centering
        \centering\includegraphics[width=0.35\columnwidth]{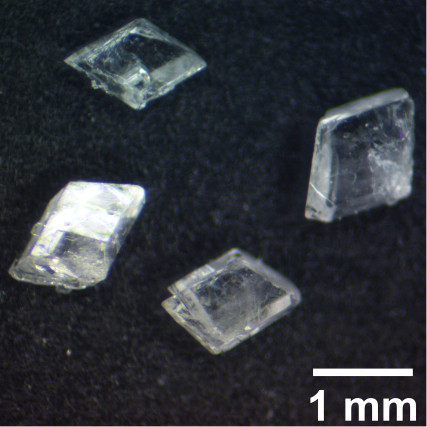}
    \caption{EF$\cdot$FA crystals with sides $\sim$1~mm}
    \label{fig:EFFAxtals}
\end{figure}

\begin{figure}
    \centering
    \begin{subfigure}[b]{0.24\columnwidth}
        \centering\includegraphics[width=\linewidth]{PL-stability-F-day0-withscale.jpg}
    \end{subfigure}
    \hfill
        \begin{subfigure}[b]{0.24\columnwidth}
        \centering\includegraphics[width=\linewidth]{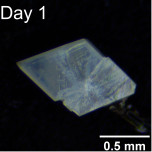}
    \end{subfigure}
    \hfill
        \begin{subfigure}[b]{0.24\columnwidth}
        \centering\includegraphics[width=\linewidth]{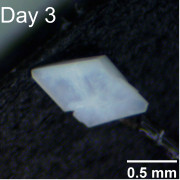}
    \end{subfigure}
    \hfill
    \begin{subfigure}[b]{0.24\columnwidth}
        \centering\includegraphics[width=\linewidth]{PL-stability-F-day7-withscale.jpg}
    \end{subfigure}
        \caption{The decomposition of a EF$\cdot$FA crystal is shown. After one day, there is visible cloudiness on the crystal as a film of EF forms on its surface. After three days, the crystal is nearly completely opaque. At this point, its room temperature photoluminescence spectrum is very close to that of pure EF. After seven days, the crystal is completely opaque.}
        \label{fig:XtalDecomp}
\end{figure}

\begin{figure}
    \centering
    \begin{subfigure}[b]{0.48\columnwidth}
        \centering\includegraphics[width=\linewidth]{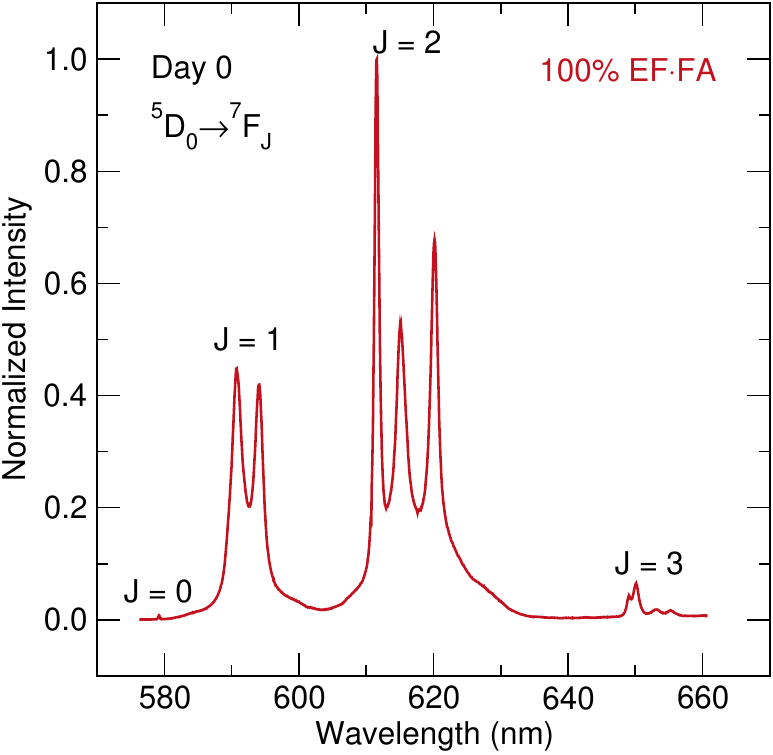}
    \end{subfigure}
    \hfill
        \begin{subfigure}[b]{0.48\columnwidth}
        \centering\includegraphics[width=\linewidth]{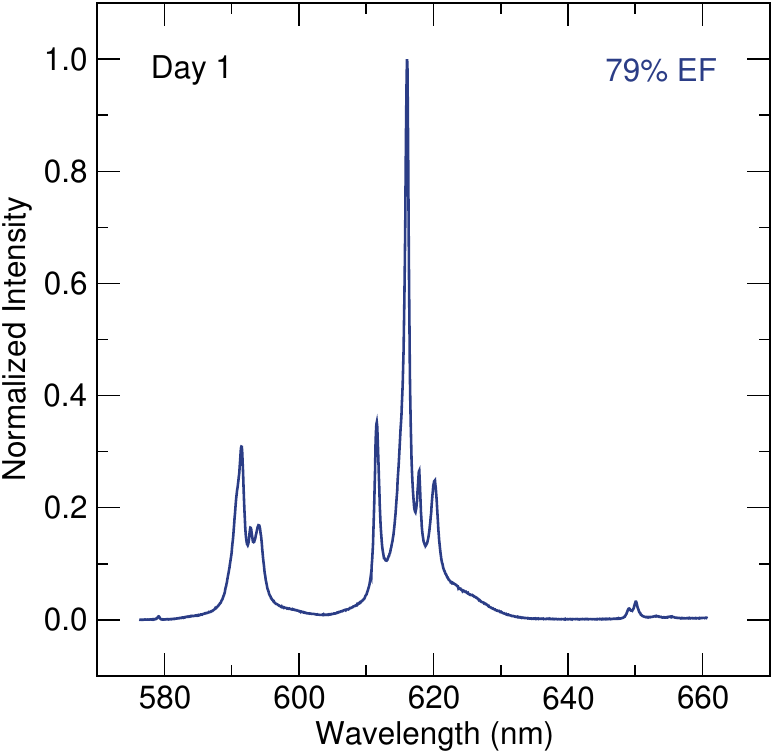}
    \end{subfigure}
    \hfill
        \begin{subfigure}[b]{0.48\columnwidth}
        \centering\includegraphics[width=\linewidth]{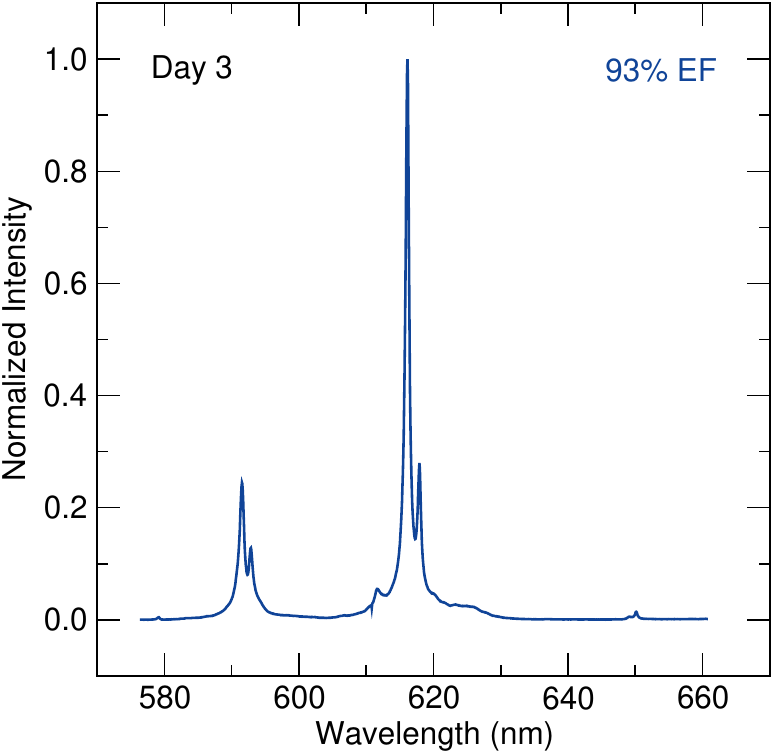}
    \end{subfigure}
    \hfill
    \begin{subfigure}[b]{0.48\columnwidth}
        \centering\includegraphics[width=\linewidth]{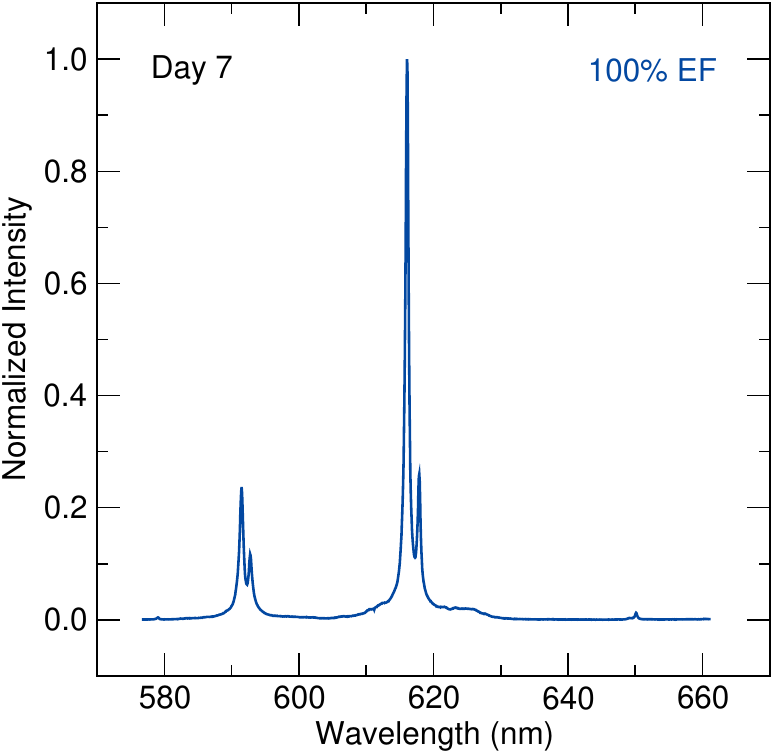}
    \end{subfigure}
        \caption{The emission spectrum of the crystal in Figure~\ref{fig:XtalDecomp} is shown for each day ($\lambda_{ex}$~=~532~nm). Overlapping of the EF and EF$\cdot$FA transitions is noticeable after only a day. Between days 3 and 7, the leftmost peak in the $J$~=~2 transition, associated with EF$\cdot$FA, disappears, and at day 7, the pattern matches closely that of pure EF. Percentages are obtained by comparing the integrated intensity of the $J$~=~2 transition to those observed before for EF and EF$\cdot$FA.}
        \label{fig:XtalDecompPL}
\end{figure}

\begin{figure}[hp]
    \centering
    \includegraphics[width=0.6\columnwidth]{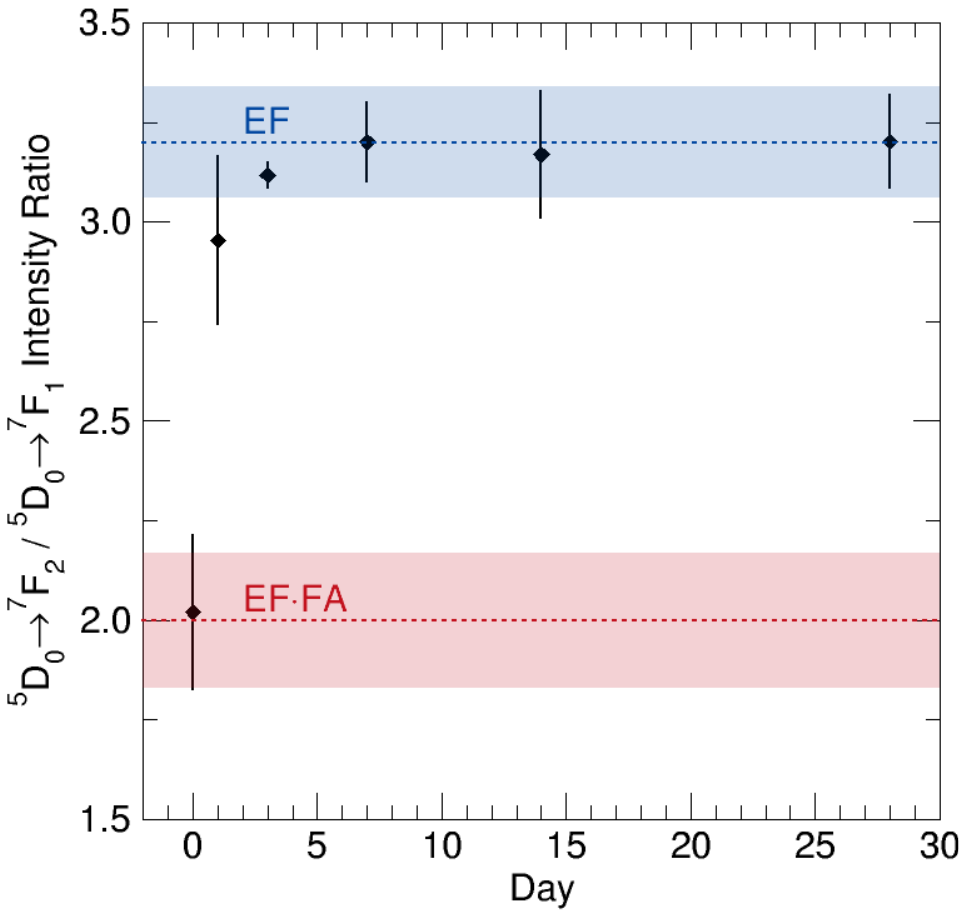}
    \caption{Transformation of the $^5$D$_0$~$\rightarrow$~$^7$F$_2$/$^5$D$_0$~$\rightarrow$~$^7$F$_1$ integrated intensity ratio of five EF$\cdot$FA crystals stored in air. Error bars indicate the sample standard deviation. The dashed lines are at the average value for EF (3.2) and EF$\cdot$FA (2.0) in a different set of crystals, and the shaded regions indicate the standard deviations associated with those averages (EF: $\pm$0.14, EF$\cdot$FA: $\pm$0.17).}
    \label{fig:PLdecomp}
\end{figure}

EF$\cdot$FA crystals begin clear (Figure~\ref{fig:EFFAxtals}) but decompose into EF in air. The significant difference in the integrated intensities of the $^5$D$_0$~$\rightarrow$~$^7$F$_2$ transition allows for an additional probe of this decomposition. Spectra were collected periodically for a set of five EF$\cdot$FA crystals. Images of the decomposition of one of these crystals are shown in Figure~\ref{fig:XtalDecomp} along with a subset of its photoluminescence emission spectrum at the same time intervals in Figure~\ref{fig:XtalDecompPL}. They were stored in air at room temperature with a relative humidity fluctuating between 50-55\%. At this humidity, the observed transitions shifted to those of EF in roughly a week (Figure~\ref{fig:PLdecomp}). 

\begin{figure}[hp]
    \centering
    \begin{subfigure}[b]{0.48\columnwidth}
        \centering\includegraphics[width=\linewidth]{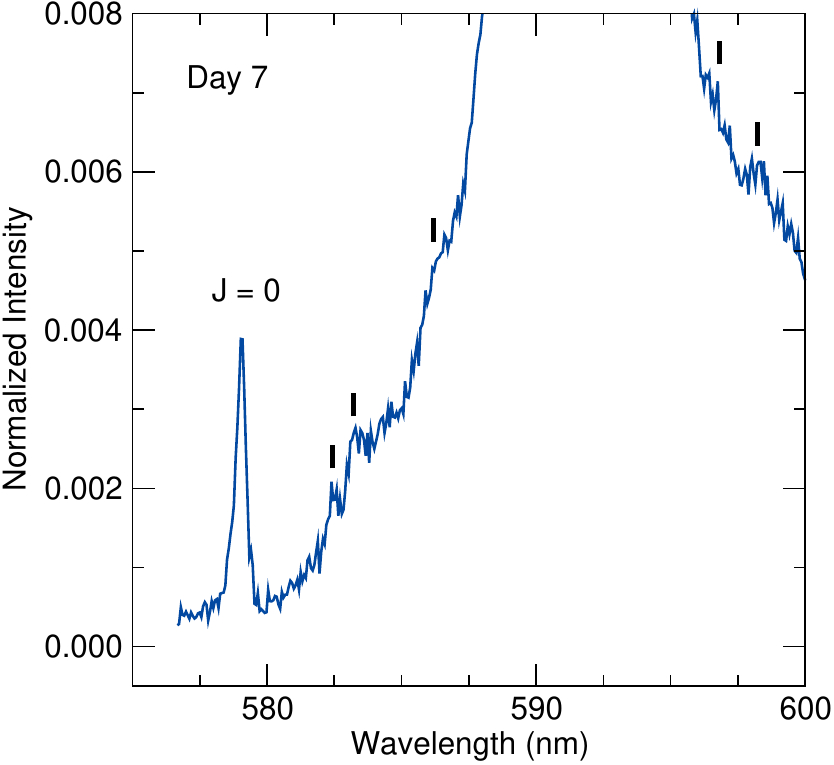}
    \end{subfigure}
    \hfill
        \begin{subfigure}[b]{0.48\columnwidth}
        \centering\includegraphics[width=\linewidth]{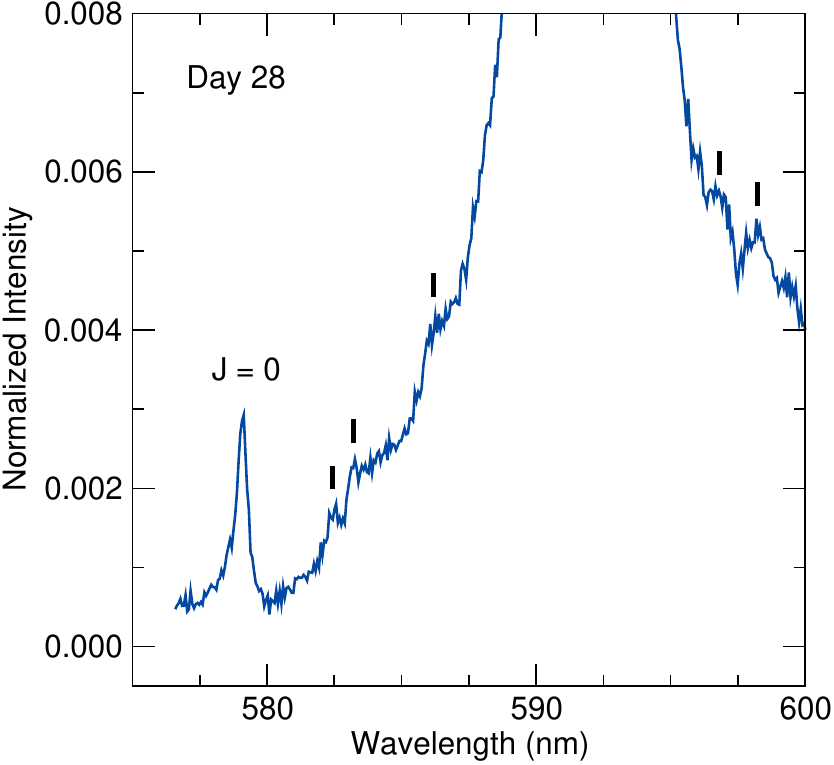}
    \end{subfigure}
        \caption{A subset of the emission spectra at days 7 and 28 for the presented decomposing crystal are shown ($\lambda_{ex}$~=~532~nm). Peak heights are normalized to the tallest $J$~=~2 peak, as before. The five additional peaks surrounding the $J$~=~1 transition are denoted with tick marks.}
        \label{fig:DecompPLsubsets}
\end{figure}

Figure~\ref{fig:DecompPLsubsets} shows that the relative height of the $J$~=~0 peak is slightly greater in the presented decomposed crystal than in the originally formed EF samples at 7 days, though it is decreasing toward the original value. These spectra also contain the five additional, low intensity peaks around the $J$~=~1 peak. After 7~days in air, the presence of the additional peaks around the $^5$D$_0$~$\rightarrow$~$^7$F$_1$ transition matches the behavior of the original, stable EF crystals.

\pagebreak 

\pagebreak

\pagebreak
 
\clearpage

\pagebreak

\begin{figure}[hp]
    \centering
    \begin{subfigure}{0.85\columnwidth}
        \centering\includegraphics[width=0.85\columnwidth]{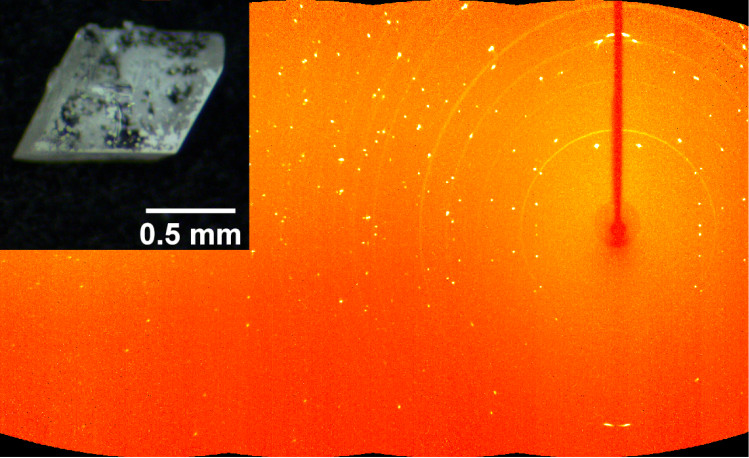}
    \end{subfigure}
    \begin{subfigure}{0.85\columnwidth}
        \centering\includegraphics[width=0.85\columnwidth]{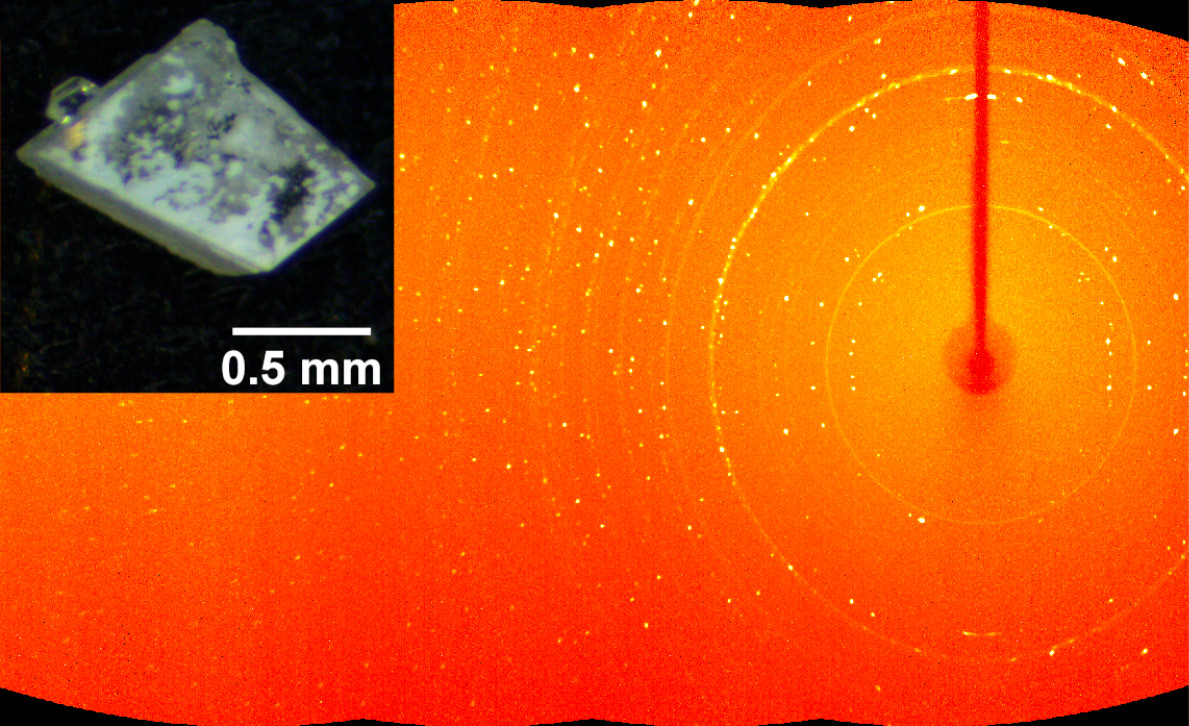}
    \end{subfigure}
        \caption{The diffraction patterns of two EF$\cdot$FA single crystals with a EF film on their surface show powder rings associated with EF and diffraction spots associated with EF$\cdot$FA.}
        \label{fig:SingleXtalDecomp}
\end{figure}

Powder diffraction patterns for decomposing EF$\cdot$FA crystals were also collected using continuous 360$^{\circ}$ rotation of the crystal in $\phi$ on a Bruker D8 Venture Duo (Cu-K$\alpha$ radiation). The crystals showed a film on their surface, and the 2D scan images contained powder rings corresponding to EF and diffraction spots corresponding to EF$\cdot$FA. Two representative crystals are shown in Figure~\ref{fig:SingleXtalDecomp}.

\pagebreak
\FloatBarrier
\begin{figure}[h]
    \centering
    \includegraphics[width=\columnwidth]{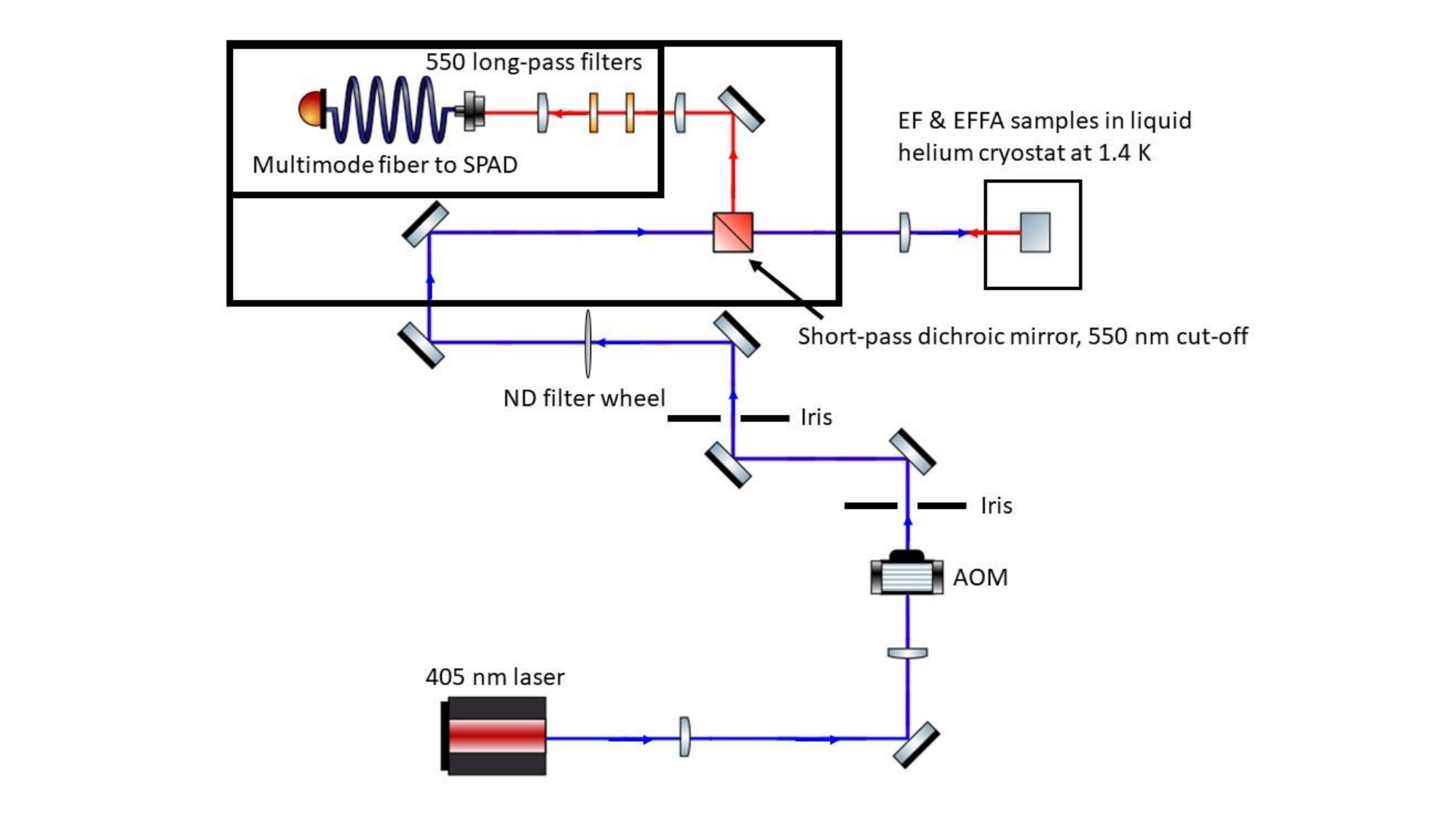}
    \caption{Optical beam path used for observing the photoluminesence decay.}
    \label{fig:beam_path}
\end{figure}
The excitation pulses for observing the photoluminescence decay were created by diffracting the light from a 405~nm excitation laser (Thorlabs CPS405) with an acousto-optic modulator (AOM) where the rf driving the AOM was switched on and off to turn the diffracted light on and off. The 405~nm excitation light was then transmitted to the EF and EF$\cdot$FA samples through a short-pass dichroic mirror with a 550~nm cut-off wavelength. A neutral-density (ND) filter after the AOM was used to attenuate the laser power. The emission from the EF and EF$\cdot$FA was reflected off the dichroic mirror, passed through a pair of 550~nm long-pass filters, and then coupled into a multimode fiber connected to a single-photon avalanche diode (SPAD) from Excelitas Technologies (model number: SPCM-AQRH-14-FC). Events from the SPAD were recorded using a Swanbian Instruments time tagger. 

\begin{figure}
    \centering
    \includegraphics[width=0.85\columnwidth]{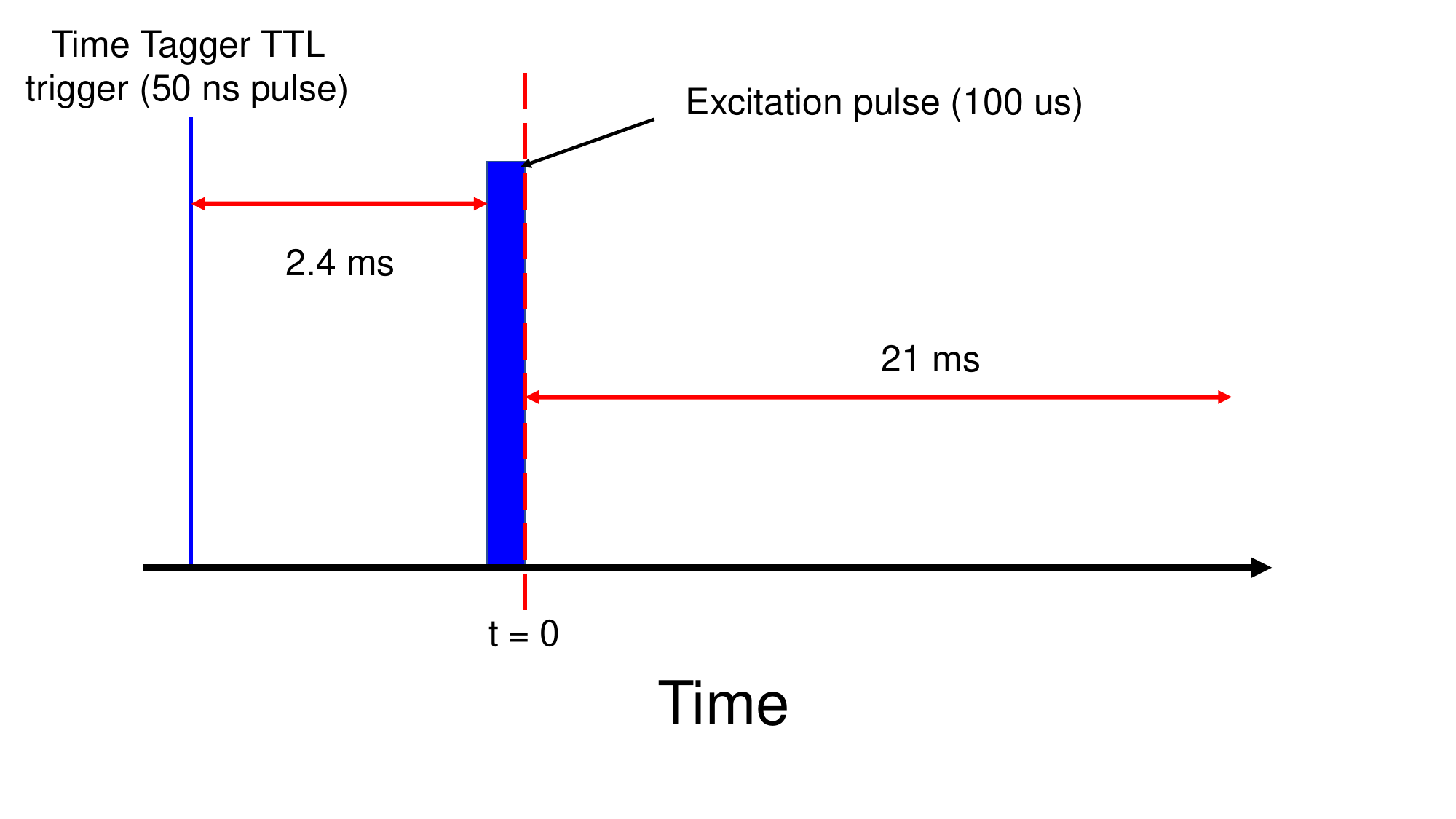}
    \caption{Example of a pulse sequence used for measuring the photoluminescence lifetime with an 100~$\mu$s excitation pulse.}
    \label{fig:pulses_sequence}
\end{figure}

The pulse sequence for the photoluminescence decay is depicted in Figure \ref{fig:pulses_sequence}. A 50 ns TTL pulse triggers the time tagger to begin recording counts from the SPAD, followed by a delay that is chosen such that the sum of the delay and the 405~nm excitation pulse duration is 2.5~ms long; then after the excitation pulse turns off there is a 21~ms time window to observe the photoluminescence decay and allow sufficient time between pulse sequences. The time tagger recorded the counts from the SPAD for 1500 time bins with 10~$\mu$s binwidths. The histogram of counts per 10~$\mu$s time bin was then averaged for 1 or 5 minutes. The room temperature and 1.4~K lifetime data were fitted to a single-exponential starting from 290~$\mu$s and 90~$\mu$s, respectively, after the excitation pulse was shut off (t = 0). There is a vertical offset in the fitting to account for the background and dark counts. Parameters for each pulse sequence used for both materials at room temperature, 1.4~K, and the R-squared value for their respective single-exponential decay fittings are listed in Table \ref{tab:lifetimes}. 
\begin{center}
\begin{table}[b]
\begin{tabular}{c|c|c|c|c|c}

    Sample & Pulse duration & Optical power & Averaging time & Lifetime & R$^{2}$ \\
    & ($\mu$s) & ($\mu$s) & (mins) & (ms) &\\
    \hline 
     EF (room temp.) & 200 & 5.96 & 1 & 1.10(1) & 0.993 \\
     EF$\cdot$FA (room temp.) & 100 & 1.36 & 1 & 1.24(1) & 0.997 \\
     EF (1.4~K) & 400 & 3.68 & 5 & 1.57(1) & 0.992 \\
     EF$\cdot$FA (1.4~K) & 300 & 7.08 & 5 & 1.43(1) & 0.987

\end{tabular}
\caption{Experimental parameters for the EF and EF$\cdot$FA photoluminescence decay at room temperature and 1.4~K, observed lifetimes extracted from the single-exponential fitting, and the R-squared from the fitting.}
\label{tab:lifetimes}
\end{table}
\end{center}